\documentclass[12pt,a4paper]{article}
\usepackage{cite}
\usepackage{graphicx}
\usepackage{amssymb}
\usepackage{epsfig}
\usepackage[tmargin=1cm,bmargin=1cm,lmargin=2cm,rmargin=2cm]{geometry}

\nopagebreak

\begin{document}
\pagestyle{empty}
\baselineskip 22pt

\begin{flushright}
SINP/TNP/2008/20\\
\end{flushright}
\vskip 65pt
\begin{center}
{\large \bf 
Diphoton signals in theories with large extra dimensions to NLO QCD at 
hadron colliders
} \\
\vspace{8mm}
{\bf
M. C. Kumar$^{a,b}$
\footnote{mc.kumar@saha.ac.in},
Prakash Mathews$^a$
\footnote{prakash.mathews@saha.ac.in},
V. Ravindran$^c$
\footnote{ravindra@mri.ernet.in},
Anurag Tripathi$^c$
\footnote{anurag@mri.ernet.in}
}\\
\end{center}
\vspace{10pt}
\begin{flushleft}
{\it
a)
Saha Institute of Nuclear Physics, 1/AF Bidhan Nagar,
Kolkata 700 064, India.\\

b)
School of Physics, University of Hyderabad, Hyderabad 500 046, India.\\

c)Regional Centre for Accelerator-based Particle Physics, 
Harish-Chandra Research Institute,
 Chhatnag Road, Jhunsi, Allahabad, India.\\
}
\end{flushleft}
\vspace{10pt}
\centerline{\bf Abstract}

We present a full next-to-leading order (NLO) QCD corrections to diphoton
production at the hadron colliders in both standard model and ADD model.  
The invariant mass and rapidity distributions of the diphotons
are obtained using a semi-analytical two cut-off phase space slicing method 
which allows for a successful numerical implementation of 
various kinematical cuts used in the experiments.  The fragmentation
photons are systematically removed using smooth-cone-isolation
cuts on the photons.
The NLO QCD corrections not only stabilise the perturbative predictions
but also enhance the production cross section   
significantly.  

\vskip12pt
\vfill
\clearpage

\setcounter{page}{1}
\pagestyle{plain}

\section{Introduction}
The gauge hierarchy problem has been one of the main motivations to go beyond
the standard model (SM).  A novel idea 
that addresses this problem was put forward by 
Arkani-Hamed, Dimopoulos and Dvali (ADD) 
wherein they introduced extra spatial dimensions and allowed 
only gravity to propagate in the extra dimensions, keeping the SM 
fields confined to a $3$-brane \cite{add}.  As the inverse square 
law behavior of gravity has so 
far been tested down to sub-millimeter length scales, the size of the extra 
dimensions, in this model, should be much smaller than sub-millimeter.
The apparent weakness of gravity as compared to the other forces 
seen in nature, can now be accounted for through the volume of the extra 
dimensions.  The relation between the fundamental scale $M_s$ at which the new physics 
sets in (above which the extra dimensions are dynamically accessible) and 
the Planck scale $M_P$ is given by
\begin{eqnarray}
M_P^2 \sim M^{(d+2)}_S~R^d ~,
\end{eqnarray} 
where $d$ is the number of extra spatial dimensions and $R$, the size of 
the extra dimensions.  Since $R$ is of order of a milli-meter, the scale 
$M_s$ can be as low as a few TeV, which circumvents the hierarchy problem.  
The propagation of a massless graviton in $4+d$ dimensions, after 
compactifying the extra dimensions on a $d$-dimensional torus, manifests 
itself as an infinite tower of massive Kaluza-Klein (KK) modes on the 
3-brane.  Each KK mode couples with SM field through energy momentum 
tensor with a coupling proportional to $\kappa \sim 1/M_P$.  However, 
the effective coupling after summing over all the KK modes is enhanced 
significantly due to large multiplicity of KK modes.  In any typical scattering
process at colliders, the gravity can enter through their KK propagator
as well as through the real emission of KK states.   These KK states are
large in number.  Hence the suppression
resulting from coupling $\kappa$ is compensated by the large multiplicity factor resulting
either from the sum of KK propagator ${\cal D}(Q^2)$ 
or from the phase space of large number of real KK states .  For example, if
the KK states enter through a propagator, we find that any typical amplitude will be
proportional to  
\begin{eqnarray}
\nonumber 
\kappa^2 {\cal D}(Q^2) &=& \kappa^2 \sum_{n} \frac{1}{Q^2-m_n^2+i\epsilon} ~,\\ 
{} &=& \frac{8\pi}{M_s^4} \left( \frac{Q}{M_s}\right)^{(d-2)}
\left[-i\pi+ 2I( \Lambda/Q) \right] ~,
\end{eqnarray}
where $\Lambda = M_s$ is the explicit cut-off on the KK sum and the function $I$
can be found in \cite{hlz}.  
Thus, for $M_s \sim {\cal{O}}$(TeV), the gravity effects can become significant and 
hence the collider phenomenology associated with this model is very 
interesting \cite{hlz}.  To exemplify, the virtual effects of the KK modes 
could lead to the enhancement of the cross sections of pair productions 
in the processes like Drell-Yan, diphoton and dijet while the 
real emissions could lead to large missing $\not{E_T}$ signals giving some new 
observable like mono-jet, mono-photon in an experiment. 
Owing to a very high
centre-of-mass energy of $\sqrt{S}=14$ TeV and 
a large gluon flux at the large hadron collider (LHC),
rich collider signals resulting from this model have been 
reported in the literature \cite{hlz,ArkaniHamed,Giudice,Eboli,Cheung,pheno}.
However, these results are based on leading order (LO) calculations.  At 
the hadron colliders like LHC, the QCD effects are often considerably large 
and hence the quantum corrections can influence the predictions significantly.
In the ADD model,  QCD effects \cite{Mathews} have been shown to increase 
the di-lepton productions and also to stabilise the perturbative predictions.
Hence in this paper we study the impact of the QCD corrections for the 
diphoton signal in the ADD model.

In QCD, the infra-red safe observable exhibit a feature called
factorisation, according to which collinear singularities 
can be factored out from the partonic cross sections in a process
independent way and then they are either absorbed 
into the bare parton distribution functions (PDF) if they originate
from initial state partons or into fragmentation functions if they 
are from final state partons.
This procedure introduces a scale called factorization scale $\mu_F$,
which is arbitrary.  In addition, ultra-violate renormalisation introduces
renormalisation scale $\mu_R$ which is again arbitrary.  The truncated 
perturbative expansion leaves our theoretical predictions $\mu_F$ and 
$\mu_R$ dependent, these scale dependence will go down as we include 
more and more terms in the expansion.  In addition, the fitted PDFs 
are usually not fully 
constrained due to insufficient experimental data.  
Hence, predictions beyond LO are often more reliable than LO ones.

Diphoton production process is an important probe for the Higgs boson 
search at the LHC.  NLO QCD corrections to this process in the SM are 
available in the literature \cite{Aurenche:1985yk,Binoth,Bern,Balazs:2007hr} 
and 
hence the diphoton signal has been a useful tool for precision studies.  
This process has also been used to search for the 
physics beyond the standard model, such as extra dimensional models, super 
symmetry and the unparticle physics.  Di-photon production \cite{Eboli} 
at Tevatron has set stringent constraints on the parameters of the ADD 
model \cite{Abazov:2008as}.  It will also play an important role at LHC.
The D\O collaboration \cite{Abazov:2008as} assumed a K-factor for their 
analysis but a full NLO QCD calculation for the ADD model does not exist 
for the diphoton production.
In this paper, we have systematically computed all the QCD effects to NLO
in perturbation theory to various important observable in 
diphoton production that are sensitive to the ADD model.  Quantitative 
estimates of QCD corrections to these observable are presented and our 
predictions are expected to be less sensitive to the factorisation scale.

\section{The Diphoton Production}
In the SM, at leading order (LO), diphoton production proceeds via
quark anti-quark annihilation subprocess 
$q+\overline q \rightarrow \gamma
+ \gamma$ \footnote {The gluon-gluon fusion process through quark loop, 
though of order $\alpha_s^2$, is comparable to the LO for studies of 
photon pairs having small invariant masses, $M_{\gamma \gamma}$. As it
falls off rapidly as $M_{\gamma \gamma}$ increases, it no longer enjoys
the status of LO process for our study on the production of large 
invariant mass photon pairs in the context of ADD model and is truly a 
NNLO contribution.}. 
In the ADD model, the SM
fields couple to KK modes through the energy-momentum tensor of the SM
fields with a strength denoted by $\kappa$.  Hence, diphotons are produced
in (i) quark antiquark annihilation ($q+\overline q 
\rightarrow \gamma + \gamma$) and (ii) gluon fusion process ($g+g
\rightarrow \gamma + \gamma$) via the exchange of KK modes.  
%
%
A comprehensive phenomenology taking into
account all the above LO processes 
has been done in \cite{Eboli}.  It was
observed that unitarity restricts the maximum value of the invariant mass
$Q$ of the diphotons.  Following \cite{Eboli}, we restrict the invariant
mass $Q$ to $Q < 0.9~M_s$.

At NLO, the SM as well as ADD leading order quark antiquark annihilation
processes get $O(\alpha_s)$ QCD radiative corrections through virtual 
gluons in 
$q +\overline q \rightarrow \gamma +\gamma + {\rm one~loop}$ and 
real gluon emissions in $ q +\overline q \rightarrow\gamma +\gamma + g$ 
processes.  
To this
order, $q (\overline q)+g \rightarrow q (\overline
q)+\gamma+\gamma $ process also shows up in both SM and ADD. The LO gluon
fusion process in the ADD model gets NLO QCD corrections to order
$\alpha_s$ through $g +g \rightarrow\gamma +\gamma + {\rm one~loop}$ and
$g +g \rightarrow\gamma +\gamma + g$ processes. Since KK modes appear at
the propagator level, the LO SM (ADD)  processes interfere with the 
corresponding NLO ADD (SM) processes giving order $\alpha_s$ NLO QCD 
corrections. 
%
We have incorporated
all these NLO QCD corrections in this article for the study that follows.

The NLO partonic cross sections are often ill-defined due to soft and
collinear singularities that result from the presence of zero momentum
gluons and mass-less partons.  In addition to these singularities, we
encounter collinear (QED)  singularities that originate when the photon in
the final state becomes collinear to the quark or the anti-quark emitting
it.  These (QED) singularities go away if we also include the diphoton
production channels resulting from the fragmentation of partons.  This
involves introduction of non-perturbative fragmentation functions. These
functions are poorly constrained. Hence, in our study we do not include
fragmentation photons but consider only direct photons. Alternatively, we
can suppress QED collinear singularities using the smooth-cone-isolation
prescription proposed by Frixione \cite{Frixione:1998jh}. In the
rapidity--azimuthal angle $(y,\phi)$ plane the amount of transverse
hadronic energy $E_T$ in any cone of radius $r=\sqrt{(\Delta y)^2 +(\Delta
\phi)^2} $ with $r < r_0$ centered around the photon must satisfy \\
%
\begin{equation}
\label{eq:isolation} 
E_T 
 \leq 
     E^{iso}_T 
     \left(
           \frac {1 - cos(r)}{1-cos(r_0)}
     \right)^n ~. 
\end{equation}
The above prescription safely removes all the photons from the 
fragmentation processes without disturbing soft and collinear partons.

An analytical computation incorporating smooth-cone-isolation 
and other kinematical constraints at NLO level is hard to achieve. 
Hence, we resort to a semi-analytical approach called
{\it two cutoff phase space slicing} method \cite{Harris:2001sx}.
In this method, two small slicing parameters $\delta_s$ and $\delta_c$ 
are introduced to isolate the cross sections that are sensitive to soft 
and collinear singularities.  The remaining part of the cross section
denoted by $d \hat \sigma^{fin}(\delta_s,\delta_c)$ is soft and collinear 
free.  The soft divergences come from virtual as well as real gluons when 
their momenta become zero.  On the other hand the collinear singularities 
arise due to mass less nature of the partons.  We compute these soft and 
collinear sensitive cross sections (they are singular in $4$ dimensions)
analytically in $ 4+ \varepsilon $ dimensions which regulate these 
singularities.  The soft singularities cancel between virtual and real 
gluons when their contributions are added appropriately.  The remaining 
collinear singular terms which appear as poles in $\varepsilon$ are 
systematically removed by collinear counter terms in $ { \overline {MS} } $ 
factorization scheme.  This is usually done at an arbitrary scale $\mu_F$.
Hence we will end up with a finite cross section coming from (a) soft and 
collinear sensitive regions denoted by $d \hat \sigma^{sc,fin}(\delta_s,
\delta_c)$ ($sc$ denotes soft and collinear) and (b) $d \hat \sigma^{fin}
(\delta_s,\delta_c)$ part of the cross section.  Their sum, ie. $(a)+(b)$,
is expected to be free of choice of the slicing parameters.   
This is an essential prerequisites for the implementation of the phase
space slicing method.

\section{Numerical Results}
In this section, we present our results for invariant mass (Q) and rapidity 
(Y) distributions of the photon pair at LHC. 
\begin{figure}[htb]
\centerline{
\epsfig{file=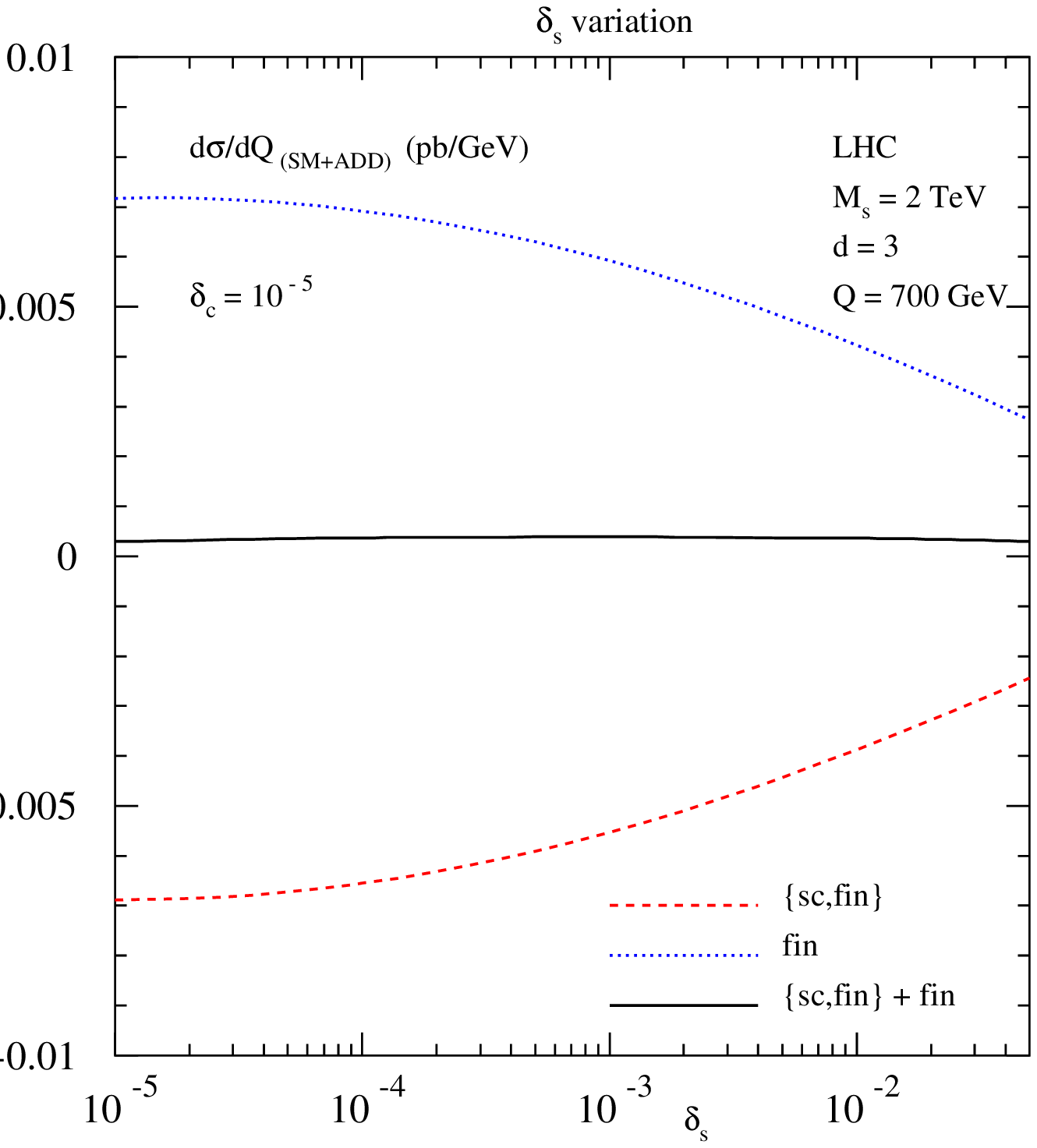,width=8cm,height=9cm,angle=0}
\epsfig{file=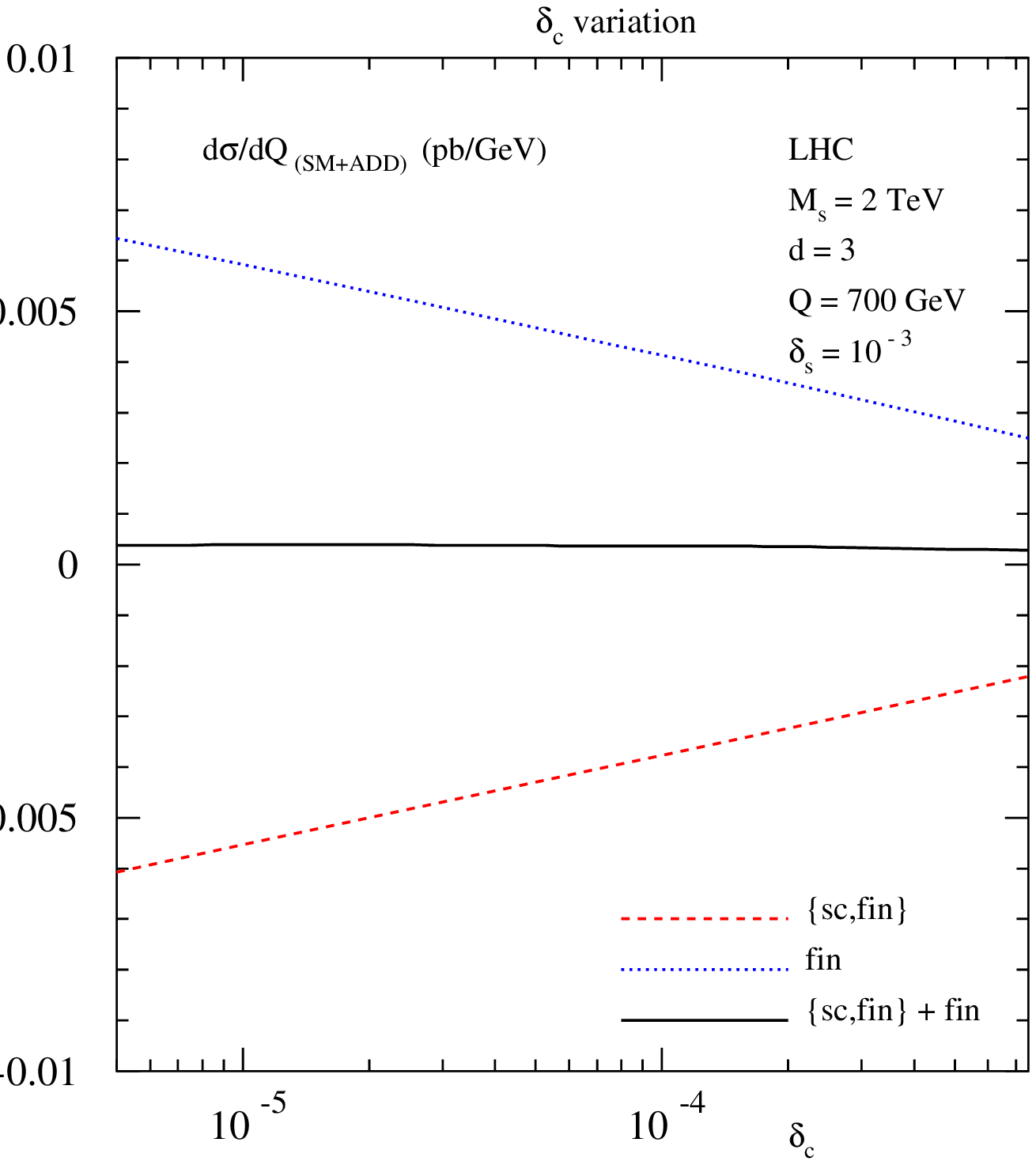,width=8cm,height=9cm,angle=0}}
\caption{Stability of the order $\alpha_s$ contribution
to the total (SM+ADD) cross section against the variation
of the phase space slicing parameters $\delta_s$ (left) and
$\delta_c$ (right) in the invariant mass distribution
of the di-photon system with $M_s=2$ TeV and  $d =3 $ at $Q=700$ GeV.}
\label{stability}
\end{figure}
We have employed the kinematical cuts given by ATLAS
collaboration \cite{atlas}: the transverse momentum
$p_T^\gamma >40~ GeV $ for the harder photons,  
$p_T^\gamma >25~ GeV $ for the softer photon, and the rapidity 
$|y_\gamma| < 2.5$ for each photon.  

In addition, the photons are isolated from hadronic
activity according to Eq.~(\ref{eq:isolation}), with 
$n=2,r_0=0.4,E^{iso}_T=15~GeV$. The minimum separation 
between the two photons is taken to be $r_{\gamma \gamma}=0.4$.
For the LO, we have used CTEQ6L PDFs and CTEQ6M for NLO
studies \cite{Pumplin:2002vw}, with the corresponding value of  
$\alpha_s(M_Z) =0.118$ and 5 light quark flavours.
The factorisation and renormalisation scales are taken to be 
$Q$, the invariant mass of the diphoton pair.
The electromagnetic coupling constant is chosen to be
$ \alpha =1/128$.  
%
%
For our numerical analysis we have chosen the ADD parameters, 
$M_s = 2$ TeV and $\Lambda=M_s$ for the number of extra spatial 
dimensions $d=3$.  This choice of $M_s=2$ is consistent with the 
limits from \cite{Abazov:2008as}.

\begin{figure}[htb]
\centerline{
\epsfig{file=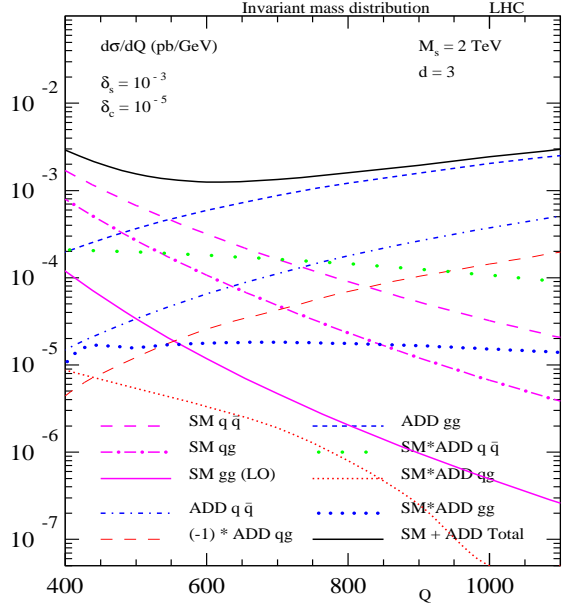,width=8cm,height=9cm,angle=0}}
\caption{Various subprocess contribution to the invariant mass distribution 
of the diphoton production with $M_s=2$ TeV and $d =3 $. The SM $gg$ 
subprocess (lower solid line) is at ${\cal O} (\alpha_s^2)$
while all other subprocess are at order ${\cal O} (\alpha_s)$.}
\label{sub}
\end{figure}
We have first checked our numerical code by studying the dependence of 
observable on the slicing parameters, $\delta_s$ and $\delta_c$.
In the left (right) panel of Fig.~\ref{stability} we have plotted the 
order $\alpha_s$ contribution to the invariant mass distribution of diphotons
in SM and ADD against the slicing parameter
$\delta_s$ ($\delta_c$) in the range between $5 \times 10^{-2}$ and $10^{-5}$.  
For the $\delta_s$ variation (left panel) we have fixed $\delta_c = 10^{-5}$
and for the $\delta_c$ variation (right panel) we have fixed $\delta_s = 10^{-3}$.
These plots show that our numerical results are least sensitive to 
the slicing parameters for a wide range.
The percentage of uncertainty that results from the choice
of slicing parameters is found to be around $6.7\%$. 
This study confirms the reliability of our code
for further predictions. 
For our numerical predictions, we have chosen
$\delta_c = 10^{-5}$ and $\delta_s = 10^{-3}$.
Other important check on our code comes from a detailed
comparison of our SM results against those given in 
the literature \cite{Aurenche:1985yk,Binoth,Bern,Balazs:2007hr}.
In particular, we find that our SM results are in very good agreement 
with those given in \cite{Bern} with their choice of parameters. 

In Fig.~\ref{sub}, we have presented various subprocess contributions to 
the invariant mass distribution of the diphoton system for the range 
$400 \leq Q \leq 1100$ GeV where gravity (through KK modes) 
contribution dominates over the SM.
Both $q\bar{q}$ and $gg$ initiated
subprocesses in ADD give large positive contributions while the $qg$
initiated subprocess gives a negative contribution.
The interference of the SM with ADD (SM*ADD) from both  $q\bar{q}$ and $gg$ subprocesses 
gives almost $Q$ independent contribution, 
while the contribution from the $qg$ subprocess falls steeply at higher values of $Q$.
Owing to the large gluon flux at the LHC, the $gg$ initiated
subprocesses in ADD give the dominant contribution 
over the rest, thus making the observable effects of ADD model clearly visible 
in the large $Q$ region.
We have also plotted the SM gluon-gluon fusion sub-process through quark loop  
contribution separately in the Fig.\ \ref{sub}.  It is clear from the plot that
its contribution is negligible compared to SM quark anti-quark initiated processes
and hence belongs to NNLO contributions.  Hence, we have not included this in our study. 

\begin{figure}[htb]
\centerline{
\epsfig{file=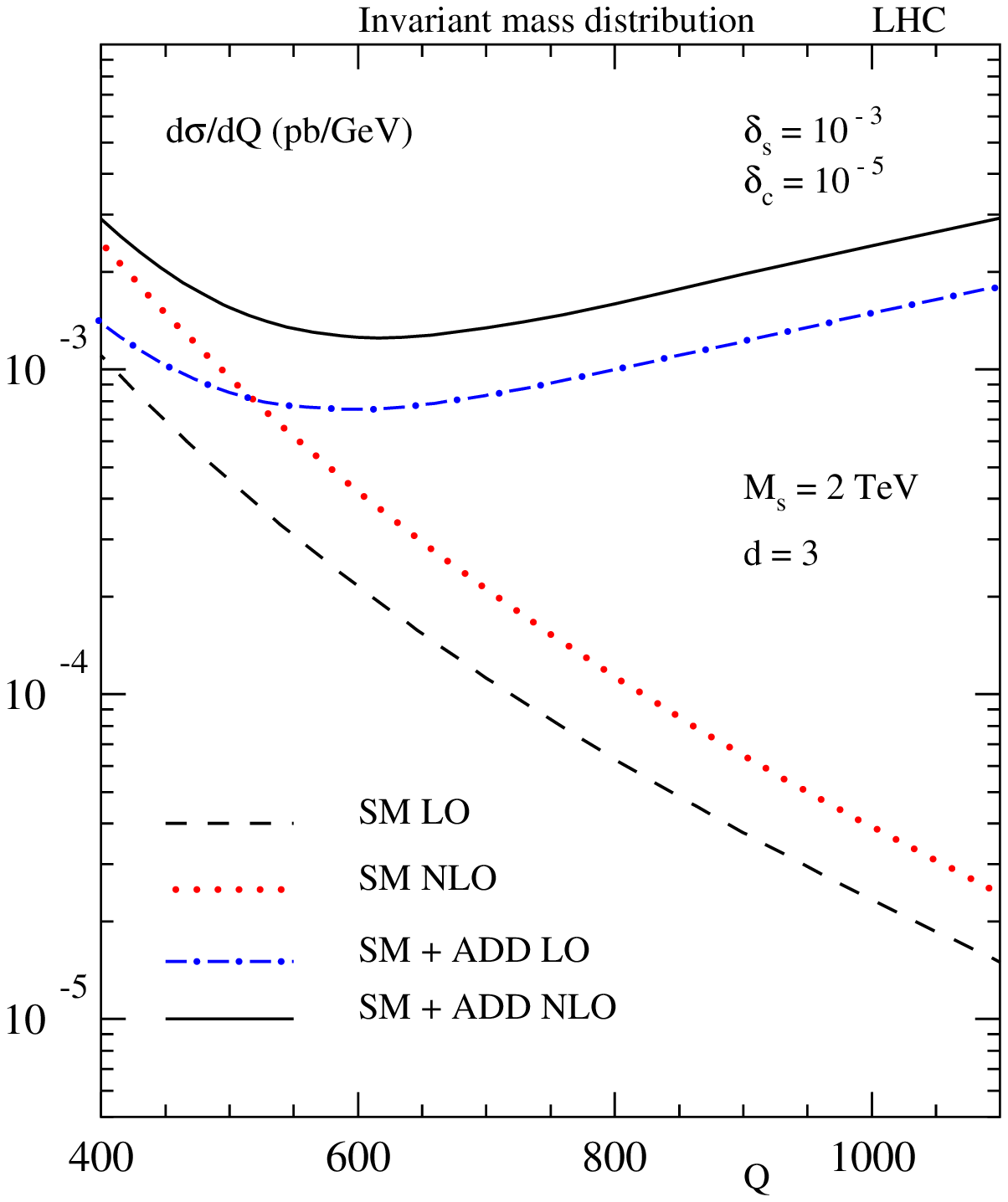,width=8cm,height=9cm,angle=0}
\epsfig{file=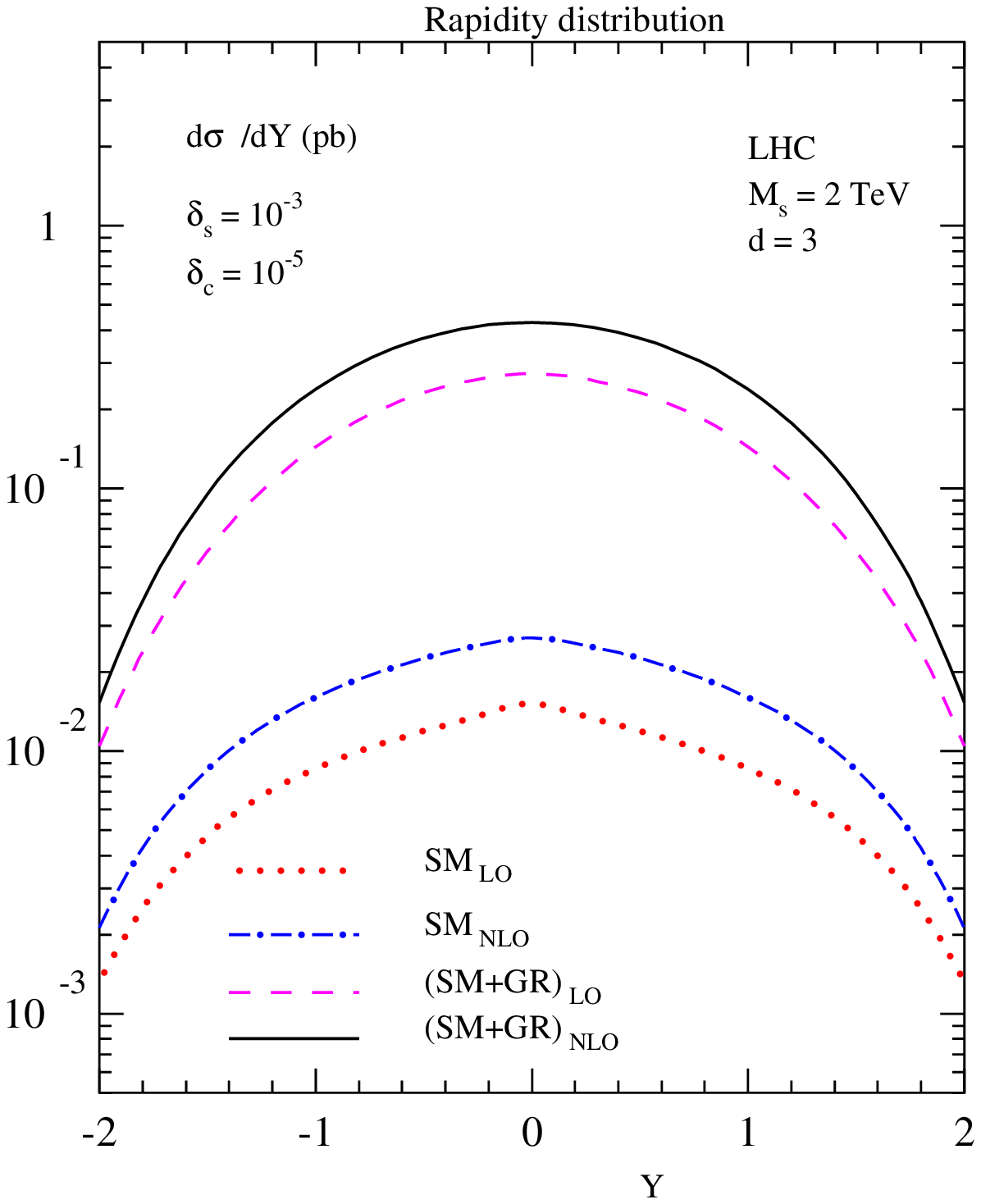,width=8cm,height=9cm,angle=0}}
\caption{Invariant mass (left) and rapidity (right) distributions
of the diphoton production at the LHC with $M_s=2$ TeV and  $d =3 $.
For rapidity distribution, we have integrated over Q in the range
$600 \leq Q \leq 1100$ GeV.}
\label{qf}
\end{figure}

\begin{figure}[htb]
\centerline{
\epsfig{file=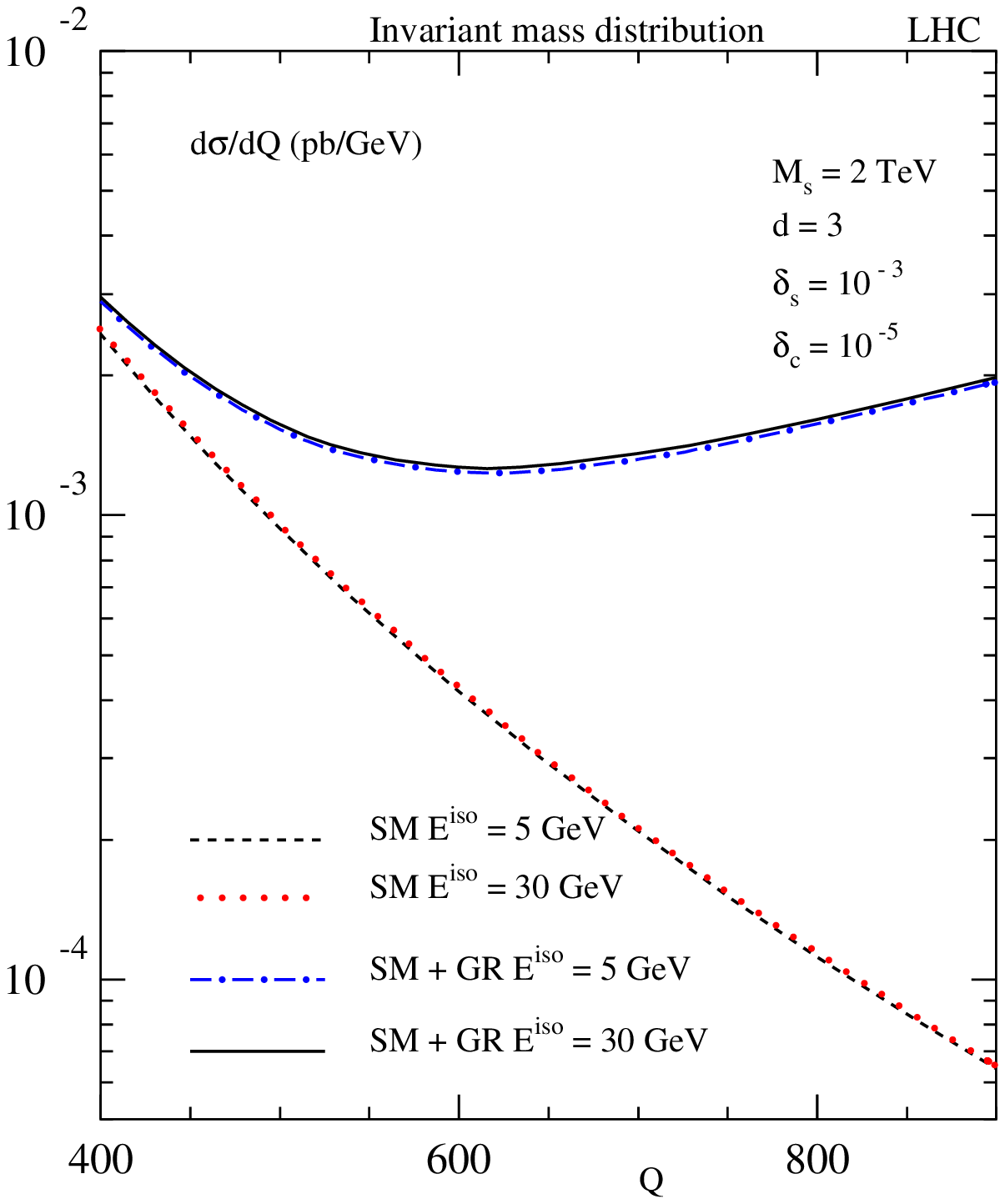,width=8cm,height=9cm,angle=0}
\epsfig{file=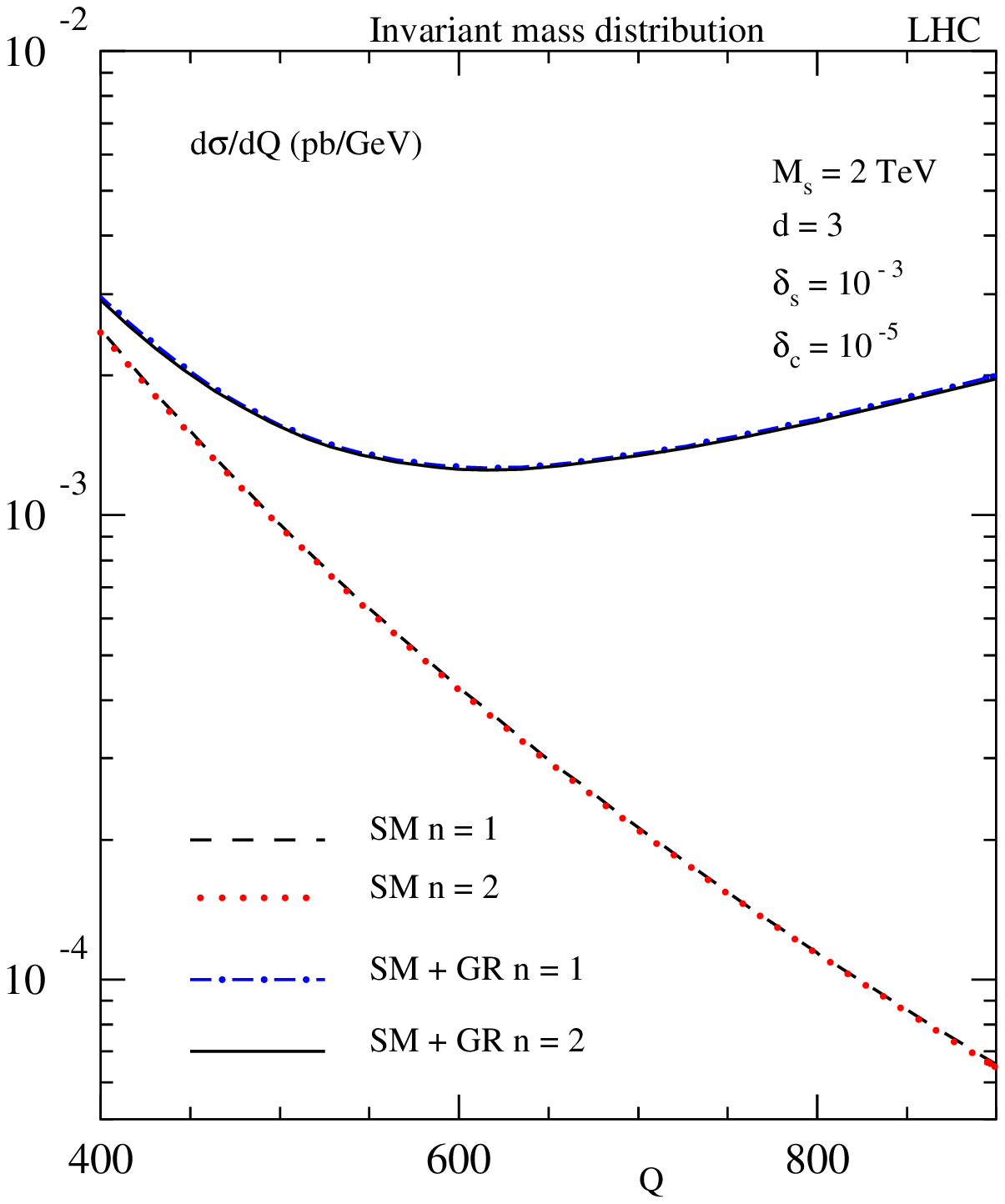,width=8cm,height=9cm,angle=0}}
\caption{
Dependence of the invariant mass distribution of the diphoton system
on the parameters $E_T^{iso}$ (left) and $n$ (right), of the Frixione's 
isolation algorithm, with $M_{s}=2$ TeV and  $d =3 $.  For the variation 
of $E_T^{iso}$ ($n$) we have kept $n$ ($E_T^{iso}$) fixed.
%
}
\label{cone}
\end{figure}

In Fig.~\ref{qf}, we have presented the invariant mass (left panel) and 
rapidity (right panel) distributions of the diphoton productions in both 
SM and ADD model.  We have plotted LO and NLO contributions separately 
to demonstrate the impact of QCD corrections.  It is clear from the plots 
that the QCD corrections to both invariant mass and rapidity distributions 
in SM as well as in ADD model are large for the entire range of $Q$ 
considered.  In the left panel we find that the contribution from ADD 
dominates over that of SM starting around $Q=500$ GeV.  The exact 
value where 
this happens depends crucially on the parameters of ADD model.  For the 
rapidity distribution (right panel), we have considered $|Y| \leq 2.0$ and 
integrated over $Q$ in the range $600 \leq Q \leq 1100$ GeV where the KK 
effects are dominant.  The cross section is found to be maximum at the 
central rapidity region both in SM and in ADD model, the later differing
by more than an order of magnitude.

\begin{figure}[htb]
\centerline{
\epsfig{file=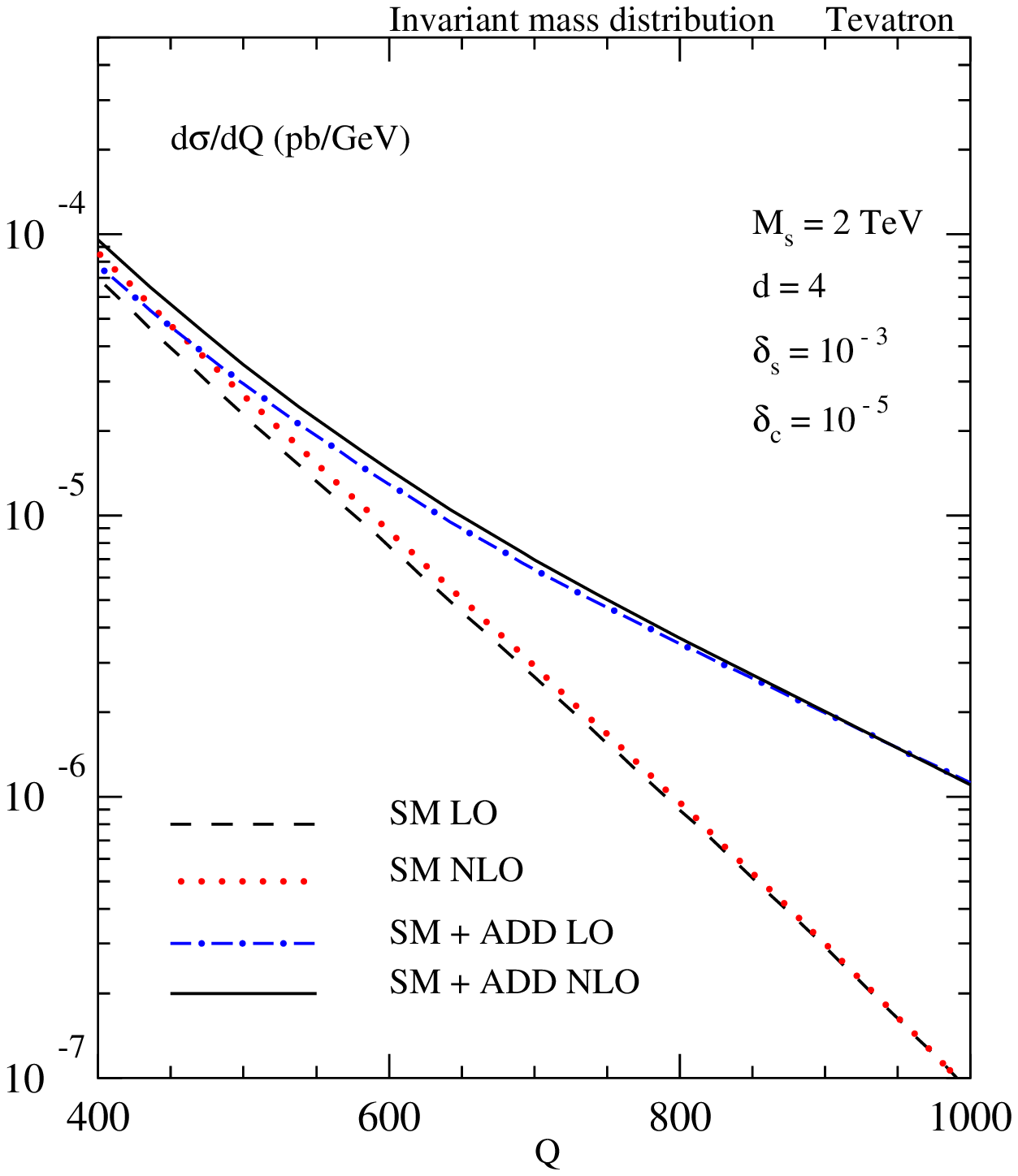,width=8cm,height=9cm,angle=0}
\epsfig{file=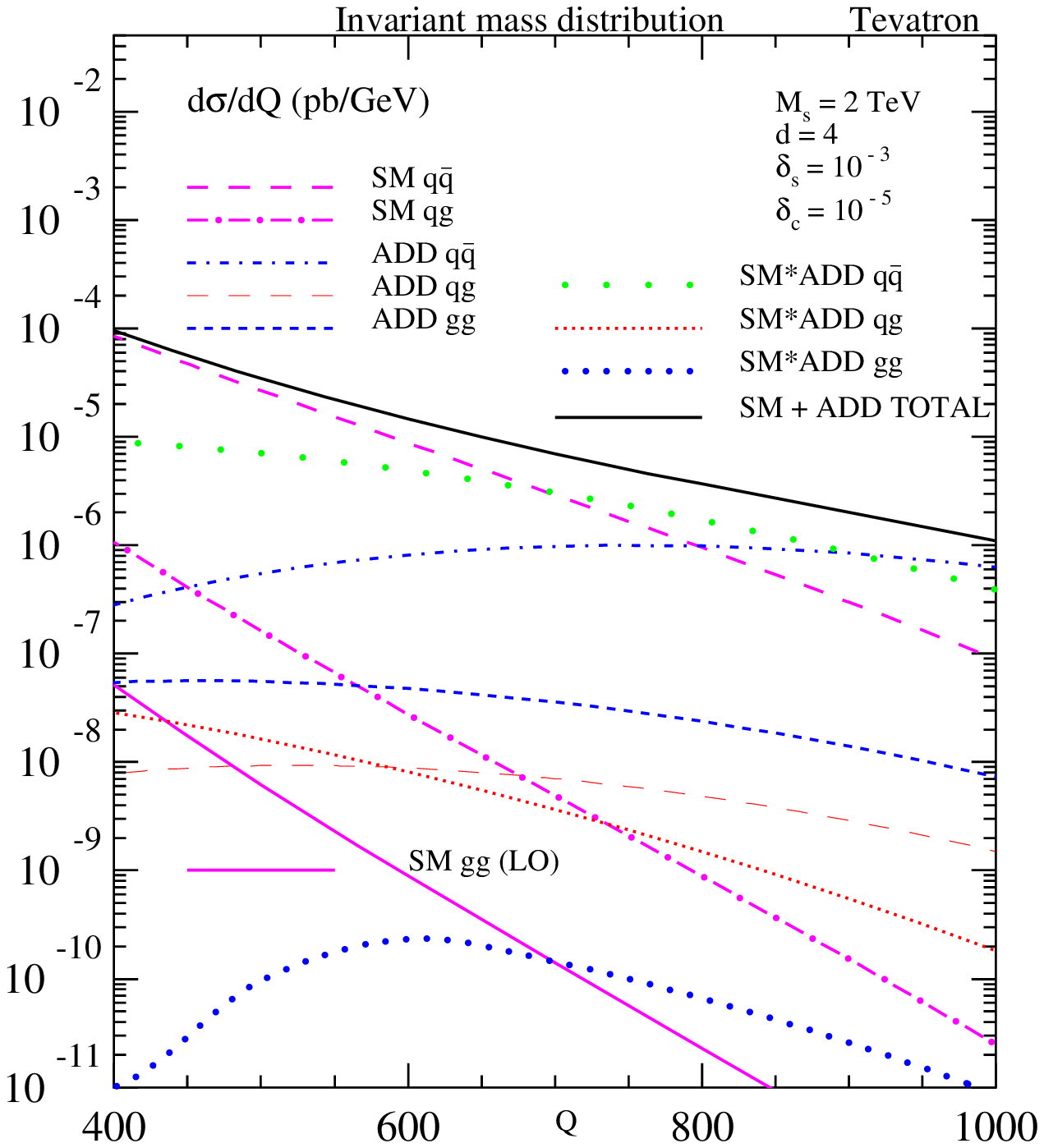,width=8cm,height=9cm,angle=0}}
\caption{Invariant mass distribution of the diphoton at the Tevatron
for $M_s=2$ TeV and  $d =4$ and in right panel the various contributing
subprocess.  The SM $gg$ subprocess is the lower solid line which is
at ${\cal O} (\alpha_s^2)$ while all other sub process is plotted
at ${\cal O} (\alpha_s)$.}
\label{tev}
\end{figure}

The cross sections do depend
on the isolation criterion.  The $E^{iso}_T$ at the partonic
level need not be the same as that of the hadrons at the detector level, 
which gives
rise to the dependency of the cross sections on $E^{iso}_T$.
In the smooth cone isolation prescription discussed above,
large logarithms of $E^{iso}_T$ often spoil the reliability of
fixed order computation.
We can study the effect the these logarithms by varying the function
that appear in the isolation criterion.  We present in Fig.~\ref{cone},
the dependency of our cross sections on 
the choices of $E^{iso}_T$ (varied between $5$ GeV and $30$ GeV), and $n$,  
(varied between $1$ and $2$).
We find that the dependency is unnoticeable 
making our predictions reliable for experimental study.


Finally we consider the invariant mass distribution at the Tevatron for
both the SM and ADD model to NLO QCD.  We have used $M_s$ value 
which is consistent with the experimental bounds \cite{Abazov:2008as} 
for the di-electromagnetic 
signal which is the combined $e^+ e^-$ and $\gamma \gamma$ final state.  
In this analysis we are hence interested only in gauging the impact of 
the QCD corrections to these studies.  In Fig.~\ref{tev} we plot the 
invariant mass distribution of the diphoton system in the range $100 < 
Q < 1000$ GeV at the Tevatron ($\sqrt{S}=1.96$ GeV) for both the SM and 
including the ADD contribution at LO and NLO in QCD.  
We have used the following kinematical cuts: (a) transverse momentum
$p_T^\gamma >15$ (14) GeV  for the harder (softer) photons, 
(b) rapidity $|y_\gamma| < 1.1$ for each photon, 
and (c) $r_0=0.4$  and $r_{\gamma \gamma}=0.4$.
In addition for the smooth-cone-isolation we use
$E_T^{iso}=2$ GeV and $n=2$.
The contributions of the various subprocess is shown in the 
right panel, for the range $400 < Q < 1000$ GeV.   We have used the 
number of extra spacial dimensions 
$d=4$ and $M_s=2$ TeV.  The impact of QCD corrections at the
Tevatron is much mild compared to the LHC where the gluonic flux is 
overwhelming.

\section{Conclusions}
In this article, we have systematically computed NLO QCD corrections to 
the diphoton production process at the hadron colliders in SM as well 
as in ADD model.  We use a semi-analytical two cut-off phase space 
slicing method to compute invariant mass as well as rapidity distributions 
of the diphotons system.  We have applied the kinematical cuts used by the 
ATLAS 
detector collaboration for our study.  A smooth-cone-isolation prescription 
on the diphotons has been used to reject poorly
known fragmentation photons.  Our method takes care of all the
soft and collinear singularities that appear at NLO level in QCD. 
We have explicitly shown that our NLO results are least sensitive to 
the slicing parameters $\delta_s$ and $\delta_c$.  Our SM results
are in good agreement with those given in the literature.
Predictions for invariant mass distribution of diphotons in ADD model 
with $M_s = 2$ TeV are found to be large compared to those in SM for 
invariant mass $Q > 600$ GeV.  This is due to large gluon flux at the 
LHC which enhances the gluon initiated production channels over the 
rest.  In addition, the QCD corrections are significantly 
large both in the SM and in the ADD over the entire range of $Q$ considered.  
For the rapidity distribution, we have integrated $Q$ in the
region $600 \leq Q \leq 1100$ GeV where the gravity (through KK modes) 
contributes significantly.
We find that the QCD corrections are important throughout the region
$|Y| \leq 2.0$.  
In addition, our results are expected to be less sensitive to the 
uncertainties coming from the choice of factorisation scale.

In summary, we have accomplished an important task of computing 
all the partonic contributions at NLO level in QCD to diphoton 
production at hadron colliders both in SM and ADD model.  
These QCD corrections for the ADD model and its interference with the
SM are being presented for the first time, while to this order
the SM results already exist in the literature.  The NLO 
QCD effects are found to be large and they are expected to reduce 
theoretical uncertainties, thus providing an excellent opportunity 
to put stringent bounds on the parameters of the ADD model when 
the experimental results are available.  Quantitative impact of the
NLO QCD corrections to both the ADD and RS model would be addressed 
in a future publication \cite{us}.

\vspace{.3cm}
\noindent
{\bf Acknowledgments:}  
MCK would like to 
thank CSIR, New Delhi for financial support.  The work of VR and AT has 
been partially supported by funds made available to the Regional Centre 
for Accelerator-based Particle Physics (RECAPP) by the Department of 
Atomic Energy, Govt.  of India.  AT and VR would like to thank the  
cluster computing facility at Harish-Chandra Research Institute where 
part of computational work for this study was carried out.



\end{document}